\documentstyle[12pt,epsfig]{article}
\begin{document}
\null\vskip1.5cm
\centerline{\Large\bf Meson electromagnetic form factors}
\vskip0.5cm
\centerline{\Large\bf in a relativistic quark model}
\vskip1.cm
\centerline{\Large L. Micu\footnote{E-mail: lmicu@theor1.ifa.ro}}

\begin{center}
{Department of Theoretical Physics\\
National Institute of Physics and Nuclear Engineering Horia 
Hulubei\\
Bucharest POB MG-6, 76900 Romania}
\end{center}
\noindent
\begin{abstract}
The main assumption of the model is that in soft processes
mesons behave like systems
made of valence quarks and an effective vacuum-like field.
The 4-momentum of the latter represents the relativistic 
generalization of the
potential energy. The electromagnetic form factors are expressed in 
terms of 
the overlap integral of the initial and final meson wave functions 
written under the form of Lorentz covariant distribution of quark
momenta. 
The calculation is fully Lorentz covariant and the form factors
of the charged mesons are normalized to unity at $t$=0.
\end{abstract}
\newpage
\vskip0.5cm
In the lowest approximation of the standard model, the elastic
electron-meson scattering is the result of the spontaneous one 
photon exchange between the electron and one of the elementary
constituents of the meson. 

The Lorentz covariant, model independent parametrization of the
hadronic matrix element entering the expression of the
scattering amplitude is: 

\begin{eqnarray}\label{def}
&&<M(P')\vert~U(+\infty,0)~J_{em}^\mu(0)~U(0,-\infty)~
\vert M(P)>\nonumber\\
&&=~{\cal T}^\mu~=~\kappa_M~f_{em}(t)~(P+P')^\mu~
\end{eqnarray}
\noindent
where $\kappa_M$ is the electric charge of the meson,
$U_s(\tau,\tau')$ is the time translation operator describing the 
evolution
of the meson under the action of strong forces, $t=(P-P')^2$ 
is the momentum transfer and the Lorentz invariant function $f_{em}(t)$ 
is the electromagnetic form factor which contains the whole
information one can obtain on the meson structure from elastic electron
scattering.

Due to the local, elementary charcater of the electromagnetic 
current $J^\mu_{em}(0)$  
the matrix element (\ref{def}) is usually related to the 
probability of finding the recoiling quark in a meson with a
momentum different from the initial one. It may be said
that the form factor shows
to what extent the initial system, where one of the quarks has been
replaced by the recoiling quark, is a meson with another momentum.
A natural consequence, strongly supported by the
experimental data, is that the form factor decreases 
with $t$, because the larger is the momentum transfer, the
harder is to incorporate the recoiling quark in a bound system.

According to this picture the calculation of the form factors resorts
to the evaluation of a kind of overlap integral and the main
problem is to find a Lorentz covariant internal wave function for a
system made out of independent constituents. This is not an easy matter, 
because as shown by the well known example of the Bethe Salpeter
equation \cite{bs} it is hard to solve this problem
without introducing some unphysical degrees of freedom.

Up to now, the most reliable results concerning the form factors 
-weak or electromagnetic-have been obtained by alternative methods, 
like, for instance, QCD sum rules \cite{ball},
lattice calculations \cite{latt}, chiral perturbation theory (CPT)
\cite{ghl}. There have also been proposed methods to calculate the 
overlap 
integral by making use of potential models \cite{isgw}, \cite{fgm},
or of some particular reference frames, where the explicit form of
the binding potential can be ignored \cite{bsw}, \cite{jaus}.

The model we use in this paper is an effective model for hadrons
as bound states of quarks, and, just like the chiral perturbation
theory \cite{hl}, it is intended to complete the low energy 
picture of QCD. Its basic features do not follow from the 
symmetry properties of the underlying theory, as in the case of
CPT, but from the general properties of the ground states.

The fundamental assumption of the model is that at low energy 
hadrons reveal a stable strucure which looks as being made
of valence quarks {\it and} of a vacuum-like field $\Phi$. The 
4-momentum carried by $\Phi$ is the relativistic generalization of 
the potential energy in the quark system. 

The main reason for introducing this effective description 
is suggested by the examples taken from the
relativistic field theory, where a bound state is not the
instantaneous effect of an elementary process, but of an infinite 
series of elementary interactions \cite{bs}. 
We conjecture therefore that the binding forces will never be
"seen" on an instantaneous picture of a bound state because they
are the result of a time average. 
Our specific assumption is that the effective component $\Phi$
represents the average over a time $T_0$ of the elementary  
quantum fluctuations generatig the binding. The time
$T_0$ depends on the underlying dynamics and must be sufficiently 
long in order to assure a stable result.

Another reason for introducing the effective field $\Phi$ 
besides the valence quarks is that a system 
made only of on-mass-shell particles having a continuous
distribution of relative momenta 
does not behave like a single particle because it does not
have a definite mass \cite{isgw}.

In agreement with these remarks, we work in momentum space where 
the mass shell constraints and the conservation laws can be easily 
expressed.

The specific assumption of the model is that the generic 
form of a single meson state is \cite{micu}:

\begin{eqnarray}
\label{meson}
&&\left.\vert M_i(P)\right\rangle=\nonumber\\
&&~{i\over(2\pi)^3}
\int~d^3~p~{m_1\over e_p}
d^3~q{m_2\over e_q}~d^4Q~ \delta^{(4)}(p+q+Q-P) \varphi(p,q;Q)
\nonumber\\
&&\times\bar u(p)\Gamma_Mv(q)~\chi^+\lambda_i\psi~\Phi^\dagger(Q)~
\left.a^\dagger(p)b^\dagger(q) 
\vert 0 \right\rangle
\end{eqnarray}

where $a^+, b^+$ are the creation operators of the valence $q\bar q$
pair; $u,v$ are Dirac spinors and $\Gamma_M$ is a Dirac matrix 
ensuring the relativistic coupling of the quark spins. The quark 
creation and annihilation operators
satisfy canonical commutation relations and commute with $\Phi^+(Q)$, 
which represents the mean result of the elementary excitations 
responsible for the binding.
Their total momentum $Q_\mu$ is not subject to any mass shell 
constraint and, in some sense, it is just what one needs to be 
added to the quark momenta in order to obtain the real meson 
momentum. This is in agreement with our assumption that $Q$ is the
relativistic generalization of the potential energy. We shall
suppose accordingly that $Q$ is time like and, from stability 
reasons, $Q_0\leq0$. 

The internal function of the meson is the Lorentz
invariant momentum distribution
function $\varphi(p,q;Q)$ which is supposed to be time independent, 
because it describes an equilibrium situation. This means that it
does not change under the action of internal strong forces and hence 
the time evolution operators $U_s(\tau,\tau')$ in eq. (\ref{def}) can 
be replaced by unity.
The main r\^ole of $\varphi$ is to ensure the single particle behaviour 
of the whole system, by cutting off the large relative momenta.
 
In the evaluation of the matrix element (\ref{def}) we shall use 
the cannonical commutation relations of the quark operators

\begin{equation}\label{cr}
\{a_i(k),a^\dagger_j(q)\}=\{b_i(k),b^\dagger_j(q)\}= (2\pi)^3
{e_k\over m}~\delta_{ij} \delta^{(3)}(k-q).
\end{equation}

and the expression of the vacuum expectation value of the effective
field which is defined as follows \cite{micu}:

\begin{eqnarray}\label{vev}
&&\left\langle 0\right\vert~\Phi(Q_1)~\Phi^+(Q_2)~\left\vert 0 
\right\rangle~=\nonumber\\
&&(VT_0)^{-1}~\int d^4~X~{\rm e}^{i~(Q_2-Q_1)_\mu~X^\mu}~=\nonumber\\
&&(2\pi)^4~(VT_0)^{-1}~\delta^{(4)}(Q_1-Q_2)
\end{eqnarray}
where $V$ is the volume of a large box and $T_0$ is the 
characteristic time involved
in the definition of the mean field $\Phi$.
It is important to remark that the definition (\ref{vev}) is 
compatible with
the norm of the vacuum state if one takes $\Phi(0)=1$. We notice
also that the relation 
(\ref{vev}) has the character of a conservation law, just like the
commutation relations (\ref{cr}), both of them being 
necessary for the fullfilement of
the overall energy momentum conservation in the process.

As a first test of the model we evaluate the norm of the
single meson state (\ref{meson}) according to the usual procedure.
The factor $\int_T~dX_0~{\rm e}^{i(E(P)-E(P'))X_0}$ coming from the 
$\delta^{(4)}$ functions in eqs. (\ref{meson}) and (\ref{vev}) shall
be put equal with $T$, because we assume that the incertitude in the
meson mass is much smaller than $T^{-1}$. A short comment concerning 
this question will be given at the end. 
Observing that $T$ is nothing else than the time involved
in the definition of the effective field $\Phi$ for a moving 
meson, we write it as $T={E\over M}~T_0$ and get:  

\begin{equation}\label{norm}
\left\langle~{\cal M}(P')~\vert {\cal
M}(P)~\right\rangle~=
2E~(2\pi)^3~\delta^{(3)}(P-P')~{\cal J}
\end{equation}
where
\begin{eqnarray} 
{\cal J}&=&{\pi\over MV}~\int d^3~p~{m_1\over
e_p}~d^3~q~{~m_2\over e_q}~d^4~Q\nonumber\\
&\times&~\delta^{(4)}(p+q+Q-P)\vert \varphi(p,q;Q)\vert^2~
Tr\left({\hat p+m_1\over 2m_1}~\gamma_5~
{\hat q-m_2\over2m_2}~\gamma_5\right)=1.
\nonumber
\end{eqnarray}

This a remarkable result because it shows that the wave function of 
the many particle 
system representing the meson can be normalized like that of a single
particle if the integral $\cal J$ converges.

As a matter of consistency, we also remark the disappearance of the
rather arbitrary time constant $T$ from the expression
(\ref{norm}) of the norm.

We evaluate now the matrix element (\ref{def}) proceeding in the 
same manner as before. By introducing the expression of the 
electromagnetic current written in terms of free quark fields
\begin{equation}
J_{em}^\mu(x)={1\over(2\pi)^3}\sum_i~\kappa_i\bar\psi_i(x)
\gamma^\mu\psi_i(x)
\end{equation}
between the meson states (\ref{meson}) and using the relations
(\ref{vev}) and (\ref{cr}) to eliminate some integrals over the
internal momenta, we obtain after a straightforward calculation:
\begin{eqnarray}\label{tmu}
&&{\cal T}_\mu~={\cal T}_\mu^{(1)}+{\cal T}_\mu^{(2)}~=\nonumber\\
&&{2\pi\over VT}~\int d^4Q~ {d^3p\over2e_p}
~{d^3q\over2e_q}~{d^3k\over2e_k}~\delta^{(4)}(p+q+Q-P)
~\varphi_i(p,q;Q)~
(t_\mu^{(1)}+t_\mu^{(2)})
\end{eqnarray}
where
\begin{eqnarray}\label{t12}
t_\mu^{(1)}&=&\kappa_1\delta^{(4)}(k+q+Q-P')~\varphi_f(k,q;Q)
\nonumber\\
&\times&Tr\left[\gamma_5(\hat k+m_1)\gamma_\mu(\hat p+m_1)
\gamma_5(-\hat q+m_2)\right]\\
t_\mu^{(2)}&=&\kappa_2~\delta^{(4)}(p+k+Q-P')~\varphi_f(p,k;Q)
\nonumber\\
&\times&Tr\left[\gamma_5(-\hat k+m_2)\gamma_\mu(-\hat q+m_2)
\gamma_5(\hat p+m_1)\right].
\end{eqnarray}

The two terms in (\ref{tmu}) represent the contributions
of the valence quarks, $k$ is the momentum of the quark
after the absorbtion of the virtual photon, $P'$ is the final meson
momentum
and $\varphi_{i,f}$ are the momentum distribution functions
of the initial and final mesons respectively. 

In the next we shall work in the reference frame where
the momenta of the initial and final mesons are
$P~=~(E,~0,~0, P)$ and $P'~=~(E,~0,~0,-P)$ respectively and the
electromagnetic form factor expresses as: 

\begin{equation}\label{ff}
f_{em}(t)~=~{1\over\kappa_M}~{1\over\sqrt{4M^2-t}}~{\cal T}_0.
\end{equation}

In this frame it is an easy matter to show that $f_{em}(0)$=1.
The demonstration makes use of  
$\delta^{(3)}(\vec{k}+\vec{q}+\vec{Q})$ to eliminate the
integrals over $\vec{k}$ 
in ${\cal T}_0^{(1)}$ and of the identity $(\hat p+m)
\gamma_0(\hat p+m)=2e_p~(\hat p+m)$ to reduce the number of 
projectors.  
Performing a similar operation on ${\cal T}_0^{(2)}$ and proceeding like
in the case of the norm, one gets
\begin{equation}
{\cal T}_0~=~2M~(\kappa_1-\kappa_2)~{\cal J}     
\end{equation}
which means 
\begin{equation}\label{qnorm}
f_{em}(0)=1
\end{equation}
if the meson wave function is properly normalized. 

In the calculation of the form factor at $t\ne 0$ we start by using
the
$\delta^{(3)}$ functions to eliminate the integrals over the momenta 
$\vec{q}$ and $\vec{k}$ in the expression of ${\cal T}_\mu^{(1)}$ and
over $\vec{p}$ and $\vec{k}$ in the expression of 
${\cal T}_\mu^{(2)}$.
After performing the traces over $\gamma$ matrices we get

\begin{eqnarray}\label{t1}
{\cal T}^{(1)}_\mu&=&{4\pi\kappa_1\over
VT}\int~de_p~dp_z~d\phi_p~d^4Q~{1\over8e_p e_k e_q}~
\delta(e_p+e_q+Q_0-E)~\delta(e_p-e_k)\nonumber\\
&\times&\varphi_i(p,q;Q)~\varphi_f(k,q;Q)
\{q_\mu~t+2\vec{P}\cdot\vec{Q}~(k_\mu-p_\mu)
\nonumber\\
&+&(k_\mu+p_\mu)[(E-Q_0)^2
+{1\over4}t-\vec{Q}^2-(m_1-m_2)^2]\}
\end{eqnarray}
and a similar result for ${\cal T}_\mu^{(2)}$.
Next, by writing 
\begin{equation}
{1\over 2e_k}\delta(e_k-e_p)={1\over4P}\delta(p_z-P)
\end{equation}
and
\begin{eqnarray}
&&{1\over2e_q}\delta(e_p+e_q+Q_0-E)=\nonumber\\
&&{1\over2p_TQ_T}\delta\left(\cos\phi_p-{(E-Q_0)^2-2e_p(E-Q_0)+
\vec{P}^2-
\vec{Q}^2+m_1^2-m_2^2\over2p_TQ_T}\right)
\end{eqnarray}
we perform the integrals over $p_z$ and $\phi_q$ in eq. (\ref{t1}).

Then the term ${\cal T}^{(1)}_0$ becomes:

\begin{eqnarray}\label{final}
&&{\cal T}^{(1)}_0={2\pi\kappa_1\over VT}~{1\over\sqrt{4M^2-t}}
\int d^4Q~\int_{e_{pm}}^{e_{pM}}de_p~\varphi_i(p,q;Q)~
\varphi_f(k,q;Q)\nonumber\\
&&\times{1\over2p_T
Q_T\sqrt{1-\cos^2\phi_p}}\{2e_p\left[(E-Q_0)^2-{1\over4}t-\vec{Q}^2
-(m_1-m_2)^2\right]\nonumber\\
&&+(E-Q_0)~t\}.
\end{eqnarray}

The integration limits over $e_p$ result from the kinematical
constraints $e_p^2\geq m_1^2+P^2$ and cos$^2\phi_p\leq 1$ 
which give:

\begin{eqnarray}\label{elim}
&&e_{pM}=
{(E-Q_0){\cal S}+
Q_T\sqrt{{\cal S}^2-4[(E-Q_0)^2-\vec{Q}_T^2](m_1^2+\vec{P}^2)}\over
2[(E-Q_0)^2-\vec{Q}_T^2]}\nonumber\\
&&e_{pm}={\rm Max}[0,{(E-Q_0){\cal S}-
Q_T\sqrt{{\cal S}^2-4[(E-Q_0)^2-\vec{Q}_T^2]
(m_1^2+\vec{P}^2)}\over2[(E-Q_0)^2-\vec{Q}_T^2]}.
\end{eqnarray}
where 
\begin{equation}
{\cal S}=(E-Q_0)^2+\vec{P}^2-\vec{Q}^2+m_1^2-m_2^2.
\nonumber
\end{equation}
The term ${\cal T}_0^{(2)}$ can be proccessed in the same manner, 
giving a similar expression.

Using the above results it is possible to calculate the
electromagnetic form factors for any momentum transfer, by choosing
an appropriate function $\varphi$.  In principle, the calculation
does not imply any other approximations, but it is hard to believe
that the multiple integral entering the expression of the form 
factor can be performed exactly.
 
The numerical results quoted in this paper have been
obtained in the
approximation $\vert\vec{Q}\vert<<\vert Q_0\vert$ in the meson
rest frame, in agreement with the assumption we made about 
the signification of $Q_\mu$.  

We used the particular Lorentz invariant distribution function 
$\varphi$ defined as
\begin{equation}\label{phi}
\varphi(p,q;Q)=N~\exp\left[-{(P\cdot
Q)^2-M^2~Q^2\over\beta^2~M^2}\right]~\exp\left[{(P\cdot Q)\over
M\alpha}\right]
\end{equation}
and performed the approximation
\begin{equation}\label{approx}
\varphi(p,q;Q)\varphi(k,q;Q)\approx\beta^3
\left({\pi\over2}\right)^{3/2} N^2 {M\over E}\delta^{(3)}(\vec{Q})
\exp\left[-{2P^2Q_0^2\over
M^2\beta^2}\right]~\exp\left[{2EQ_0\over M\alpha}\right] 
\end{equation}
expected to be valid for a small parameter $\beta$.

The approximation (\ref{approx}) allows one to do immediately 
the integration over $\vec{Q}$ and $e_{p}$ in ${\cal T}_0^{(1)}$
leaving only the integral over $Q_0$ to be performed.
The integration limits (\ref{elim}) generated by the kinematical
constraints become now:
\begin{eqnarray}\label{newc}
e_p^2&=&\left[{(E-Q_0)^2+\vec{P}^2+m_1^2-m_2^2\over2(E-Q_0)}\right]^2
\geq m_1^2+\vec{P}^2\\
e_q&=&{(E-Q_0)^2+\vec{P}^2-m_1^2+m_2^2\over2(E-Q_0)}\geq m_2,
\end{eqnarray}
leading to a single condition for the integral over $Q_0$ in 
${\cal T}_0^{(1)}$, namely:
\begin{equation}
Q_0~\leq Q^{(1)}_{0M}={\rm Min}[0,\sqrt{M^2+\vec{P}^2}-m_2-
\sqrt{\vec{P}^2+m_1^2}].
\end{equation}

Performing the same operation in ${\cal T}_0^{(2)}$ and using the 
normalization condition (\ref{norm}) to eliminate the constant 
$N$, we finally get:
\begin{eqnarray}\label{fin}
&&f_{em}(t)={\pi\over {\cal J}PT}~{M^2\over E^2}
\nonumber\\
&&\times\left\{\kappa_1\int_{-\infty}^{Q_{0M1}}dQ_0~{\rm exp}\left(
{tQ_0^2\over 4M^2\beta^2}\right)\exp\left[{2EQ_0\over M\beta}\right]
{1\over(E-Q_0)}\right.
\nonumber\\
&&\times\left[2e_p\left((E-Q_0)^2-{1\over4}t-(m_1-
m_2)^2\right)+t(E-Q_0)\right]\nonumber\\
&&-\kappa_2\int_{-\infty}^{Q_{0M2}}dQ_0~{\rm
exp}\left[{tQ_0^2\over 4M^2\beta^2}\right]
\exp\left[{2EQ_0\over M\beta}\right]
{1\over(E-Q_0)}\nonumber\\
&&\left.\times\left[2e_q\left((E-Q_0)^2-
{1\over4}t-(m_1-m_2)^2\right)+t(E-Q_0)\right]\right\},
\end{eqnarray} 
where $t=-4\vec{P}^2,~e_{p,q}={(E-Q_0)^2-{1\over4}t\pm m_1^2\mp
m_2^2\over2(E-Q_0)}$ and 
\begin{equation}
{\cal J}=\int_{-\infty}^0
dQ_0\left[1-{(m_1-m_2)^2\over(M-Q_0)^2}\right]~\exp\left({2Q_0\over
\alpha}\right)
\sqrt{\left[(M-Q_0)^2-m_1^2-m_2^2\right]^2-4m_1^2m_2^2}.
\nonumber
\end{equation}

It is easy to see that $f_{em}^{\pi^0,\eta,\eta'}(t)\equiv0$,
while $f_{em}^{K^0,\tilde K^0}(t)\sim (m_s-m_d)$ and in principle
it does not vanish. 

The expression (\ref{fin}) is, of course, valid for $t\ne0$, but the
infinite value one gets in the limit $t\to0$ seems to contradict
the normalization of the electric charge (\ref{qnorm}) which has been
demonstrated previously.

This is a disturbing question which deserves a careful examination.
Looking back, we remark that the contradiction comes from the evaluation
of some $\delta$ functions:
\begin{eqnarray}
&&\delta^{(3)}(\vec{p}+\vec{q}+\vec{Q}-\vec{P})\delta(e_p+e_q+Q_0-
E(P))\nonumber\\
&&\times\delta^{(3)}(\vec{k}+\vec{q}+\vec{Q}'-\vec{P}')\delta(e_k+
e_q+Q_0-E(P'))\delta^{(4)}(Q-Q')
\end{eqnarray}
which have been written as
\begin{eqnarray}\label{td0}
&&\delta^{(3)}(\vec{p}+\vec{q}+\vec{Q}-\vec{P})\delta(e_p+e_q+Q_0-
E(P))\nonumber\\
&&\times\delta^{(3)}(\vec{p}-\vec{k}-2\vec{P})\delta(e_p-e_k)
\delta^{(3)}(\vec{Q}-\vec{Q}')\delta(Q_0-Q'_0)
\end{eqnarray}
at $t\ne$0, while at $t$=0 they have been written as
\begin{eqnarray}\label{te0}
&&\delta^{(3)}(\vec{p}+\vec{q}+\vec{Q})\delta(e_p+e_q+Q_0-M)
\nonumber\\
&&\times\delta^{(3)}(\vec{p}-\vec{k})\delta(Q_0-Q'_0-M+M')\delta^{(3)}
(\vec{Q}-\vec{Q}'){1\over2\pi}\int {\rm e}^{i(M-M')X_0} dX_0
\end{eqnarray}
and the integral has been replaced by $T$ because it was assumed 
that the uncertainty in the meson mass is much smaller than 
$T^{-1}$.

The problem comes from the fact that $T$ is finite and hence
it is illegal to put $\delta(Q_0-Q'_0)$ in the expression of the
vacuum expectation value (\ref{vev}). This means that instead
of (\ref{td0}) we ought to write 
\begin{eqnarray}
&&\delta(e_p+e_q+Q_0-E(P))\delta(e_p-e_k+Q_0-Q'_0-E(P)+E'(P'))
\nonumber\\
&&\times\delta^{(3)}(\vec{p}+\vec{q}+\vec{Q}-\vec{P})
\delta^{(3)}(\vec{p}-\vec{k}-2\vec{P})
{1\over2\pi}\int~{\rm e}^{i(E(P)-E'(P')-e_p+e_k)X_0}
dX_0
\end{eqnarray}
and perform the calculation with $T$ finite by also taking into
account the indetermination of the meson mass.
In the present calculation we do not follow this line
because it is very cumbersome. Instead of this, we use the charge
normalization condition (\ref{qnorm}) in order to fix the parameter 
$T$, which is mainly the same thing. 

To this end we notice that $T$ is the overlapping time of the 
complex systems representing the initial and final mesons. Then, as 
resulting from a careful analysis of the relations (\ref{td0}) and
(\ref{te0}), one must write $T={T_0\over v}$ where
$v=\sqrt{{(P\cdot P')^2 \over M^4}-1}={2PE\over M^2}$ is the 
relative velocity of the two mesons. This solves the problem and
the limit $t\to 0$ can now be freely performed in eq.(\ref{fin}).  

By using different values for the cut-off parameters $\alpha$ and
$\beta$ 
we found that the charge radii increase with $\alpha$ and decrease
when the parameter $\beta$ increases.
We also found that the shape of the electromagnetic
form factor $f(t)$ depends on the ratio $\rho={\alpha^2\over M\beta}$.
For $\rho>1$ the shape is exponential, leading to large values for 
the charge radius, while for $\rho<1$ it changes 
and the radius can be as small as wanted.

The dependence of the shape on the ratio $\rho$ is
illustrated by the plots of the pion form factor in Fig.1.

\begin{figure}[htb]
\begin{center}
\epsfig{figure=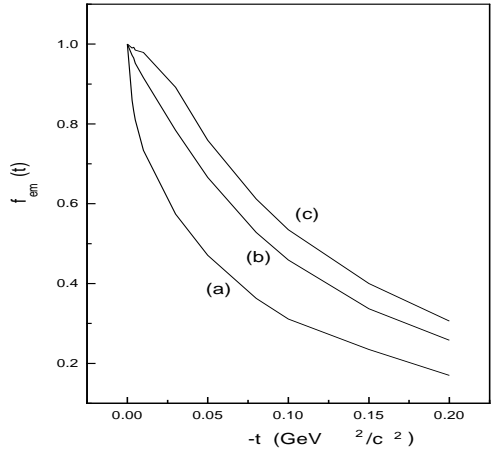,width=70mm,height=75mm}
\caption
{Plots of the pion electromagnetic form factor for the 
following values of the parameters:  
$\alpha=0.3~M_\pi$; (a) $\beta$=0.02 GeV; (b) $\beta$=0.04
GeV; (c) $\beta$=0.06 GeV.}
\end{center}
\end{figure}

For the comparison with the experimental data we shall retain 
however only
the cases with $\rho \ge1$, which fit 
the expected growth of the form factor at time-like $t$ and are 
also in agreement with our initial assumption $Q^2>0$. 

The quark masses used in the calculations 
have been determined together with the cut-off parameter
$\alpha$ from the fit of the decay constants of pseudoscalar
mesons \cite{micu}. We take:
$m_u=7$MeV, $m_d=$10MeV, $m_s$=
400MeV and $\alpha=0.3~M$, which give $F_\pi:F_K:F_D:F_B:F_{D_s}=
130.:160:254:144:386.$, in agreement with the experimental data.
We note that the values of the quark masses are rather close  
to the values suggested by chiral symmetry scheme \cite{leut}.

Moreover, 
by using the normalization condition (\ref{qnorm}) we get in the
charged pion case $T_0\approx10^{-22}$s, which is in agreement
with the low values for the quark masses we used. 

Taking now $\rho\approx1. (\beta=0.04$ GeV in the pion case), we
find the following values for the charge radii: $r_{\pi^\pm}^2=
1.6$ fm$^2$, $r_{K^\pm}^2$=1.0 fm$^2$, $r_{K^0}$=-0.17 fm$^2$, which 
are much larger than the
measured ones: $r_{\pi^\pm}$=0.44 fm$^2$, $r_{K^\pm}$=0.29 fm$^2$,
$r_{K^0}$=-0.054 fm$^2$ \cite{data}.
We notice, however, the negative sign of $r_{K^0}^2$ in agreement
with the experimental result.This shows that the contribution of the
heavy quark is dominant.

The values we obtained for the charge radii suggest that the
approximation (\ref{approx}) is inadequate. A simple way to improve 
it is to replace 
the symmetry scheme based on the full Lorentz group with the
symmetry under the collinear group which is equivalent with the 
flux tube model with frozen transverse degrees \cite{mb}.
The longitudinal and the temporal degrees of 
freedom are then the only active and the multiple integral
in eq.(\ref{final}) reduces to a simple one. 
By using less drastic cuts of the internal momenta we expect to 
obtain a slower decrease of the
form factors and a better agreement with the experimental data.
The work on this line is in progress.

\vskip0.5cm
{\bf Acknowledgements} 

The author thanks Prof. H. Leutwyler for the kind hospitality 
at the Institute of Theoretical Physics of the University of Bern.
This work was completed during author's visit at ITP-Bern,
which has been supported by 
the Swiss National Science Foundation  under Contract No. 7 IP 051219. 
The partial support of the Romanian Academy through the 
Grant No. 329/1997 is also acknowledged.


\vskip0.2cm
\noindent
\begin{thebibliography}{99}
\bibitem{bs}
E. E. Salpeter and H. A. Bethe Phys. Rev. 84 (1951) 1232; M. Gell-Mann
and F. Low, Phys. Rev. 84 (1951) 350; N. Nakanishi, Prog. Theor. 
Phys. Suppl. 43 (1969) 1.
\bibitem{ball}
M. Neubert, Phys. Rep. 245 (1994) 259.
\bibitem{latt}
R. C. Brower, M. B. Gavela, R. Gupta and G. Maturana, Phys. Rev. Lett.
53 (1984) 1318; N. Cabibbo, G. Martinelli and R. Petronzio Nucl. Phys.
B 244 (1984) 381.
\bibitem{ghl}
J. Gasser and H. Leutwyler, Nucl. Phys. B 250 (1985) 465; 517; 539.
\bibitem{isgw}
B. Grinstein, M. B. Wise and N. Isgur, Phys. Rev. Lett. 56 (1986) 298;
N. Isgur, D. Scora, B. Grinstein and M. B. Wise, Phys. Rev. D 39 
(1989) 799.
\bibitem{fgm}
R. N. Faustov, Ann. Phys. (N.Y.) 78 (1973) 176.
\bibitem{bsw}
M. Wirbel, B. Stech and M. Bauer, Z. Phys. C 29 (1985) 637.
\bibitem{jaus}
P. L. Chung, F. Coester and W. N. Polizou, Phys. Lett. B 205 (1988) 
545; W. Jaus, Phys. Rev. D 41 (1990) 3394.
\bibitem{hl}
H. Leutwyler, Ann. Phys. (N.Y.) 235 (1994) 165. 
\bibitem{micu} 
L. Micu,  Phys. Rev. D 55 (1997) 4151. 
\bibitem{leut}
H. Leutwyler, Phys. Lett. B378 (1996) 313.
\bibitem{data}
S. R. Amendolia et al. (NA7 coll.) Nucl. Phys. B277 (1986) 168;
W. R. Molzen et al. Phys. Rev. Lett. 41 (1978) 1213.
\bibitem{mb}
M. Burkhardt, Phys. Rev. D 56 (1997) 7501.
\end{thebibliography}
\end{document}